\documentclass[12pt,usenames,dvipsnames,a4paper]{article}

\usepackage{soul}

\usepackage[utf8]{inputenc}
\usepackage[T1]{fontenc}

\usepackage{changepage}
\usepackage{amsmath,amsfonts,amssymb}
\usepackage{epsf,amsmath,bbold,amsfonts,stmaryrd}
\usepackage{mathrsfs}
\usepackage{appendix}
\usepackage{caption}
\usepackage{color}
\usepackage{datetime}
\usepackage{float}
\usepackage{graphicx}
\usepackage[colorlinks,hyperindex,breaklinks]{hyperref}
\hypersetup{pageanchor=false,citecolor=red,urlcolor=red}
\usepackage{indentfirst}
\usepackage[numbers,square,comma,sort&compress,merge]{natbib} 
\usepackage{subcaption}
\usepackage{mathtools}
\usepackage{ytableau}
\usepackage{tabu}
\usepackage{esvect}
\usepackage{physics}

\usepackage{calc}
\usepackage{bm}

\allowdisplaybreaks

\def\a{\alpha}
\def\arctanh{\mathrm{arctanh}}

\def\c{\chi}

\def\d{\delta}

\def\eps{\varepsilon}
\def\f{\frac}
\def\g{\gamma}

\def\G{\Gamma}

\def\l{\left}

\def\mc{\mathcal}

\def\m{\mu}
\def\n{\nu}

\def\p{\partial}

\def\r{\right}
\def\s{\sigma}
\def\t{\tau}

\def\x{\xi}

\def\z{\zeta}

\def\be{\begin{equation}}
\def\ee{\end{equation}}

\def\bea{\begin{eqnarray}}
\def\eea{\end{eqnarray}}

\def\ba{\begin{array}}
\def\ea{\end{array}}

\def\bc{\begin{center}}
\def\ec{\end{center}}

\def\bl{\begin{flushleft}}
\def\el{\end{flushleft}}

\def\br{\begin{flushright}}
\def\er{\end{flushright}}

\def\bi{\begin{itemize}}
\def\ei{\end{itemize}}

\def\bt{\begin{tabular}}
\def\et{\end{tabular}}

\makeatletter

\newsavebox\myboxA
\newsavebox\myboxB
\newlength\mylenA

\newcommand*\xoverline[2][0.75]{%
\sbox{\myboxA}{$\m@th#2$}%
\setbox\myboxB\null% Phantom box
\ht\myboxB=\ht\myboxA%
\dp\myboxB=\dp\myboxA%
\wd\myboxB=#1\wd\myboxA% Scale phantom
\sbox\myboxB{$\m@th\overline{\copy\myboxB}$}%  Overlined phantom
\setlength\mylenA{\the\wd\myboxA}%   calc width diff
\addtolength\mylenA{-\the\wd\myboxB}%
\ifdim\wd\myboxB<\wd\myboxA%
\rlap{\hskip 0.5\mylenA\usebox\myboxB}{\usebox\myboxA}%
\else
\hskip -0.5\mylenA\rlap{\usebox\myboxA}
{\hskip 0.5\mylenA\usebox\myboxB}%
\fi}
\makeatother

\def\be{\begin{equation}}
\def\ee{\end{equation}}

\def\bea{\begin{eqnarray}}
\def\eea{\end{eqnarray}}

\def\f{\frac}

\def\p{\partial}

\usepackage{xspace}

\newcommand*{\eq}{eq.\@\xspace}
\newcommand*{\eqs}{eqs.\@\xspace}

\usepackage{xifthen}% Provides \isempty test
\newcommand*\diff{\mathrm{d}} % Straight differential
\newcommand*\ldiff[2][]{ \ifthenelse{\isempty{#1}}{ \diff
#2}{\diff^#1#2} \,} % Differential with measure; the mandatory argument
\let\limitint\int % Only when I provide explicit limits for the
\renewcommand{\int}{\limitint \!} % The standard integral should have
\begin{document}

\begin{titlepage}

\vspace*{-1cm}

\begin{adjustwidth}{-1.3cm}{-.7cm}

\begin{center} 
\Large\textbf{Weyl-invariant Einstein-Cartan gravity with a heavy ALP: \\
 Higgs Inflation and $\a$-attractors }
\end{center}

\end{adjustwidth}

\begin{center}
\textsc{Georgios K. Karananas,$^\star$
~Mikhail Shaposhnikov,$^\dagger$
~Sebastian Zell\,$^{\star,\ddagger}$}
\end{center}

\begin{center}
\it{$^\star$Arnold Sommerfeld Center\\
Ludwig-Maximilians-Universit\"at M\"unchen\\
Theresienstra{\ss}e 37, 80333 M\"unchen, Germany\\
\vspace{.4cm}
$^\dagger$Institute of Physics \\
\'Ecole Polytechnique F\'ed\'erale de Lausanne (EPFL) \\ 
CH-1015 Lausanne, Switzerland\\
\vspace{.4cm}
$^\ddagger$Max-Planck-Institut f\"ur Physik\\
Boltzmannstr. 8, 85748 Garching b. M\"unchen, Germany}
\end{center}

\begin{center}
\small
\texttt{\small georgios.karananas@physik.uni-muenchen.de}  \\
\texttt{\small mikhail.shaposhnikov@epfl.ch} \\
\texttt{\small sebastian.zell@lmu.de} 
\end{center}

\begin{abstract} 

We initiate the analysis of the inflationary dynamics in Weyl-invariant
Einstein-Cartan gravity nonminimally coupled to the Standard Model of
particle physics. We take the axion-like particle of gravitational origin to
be heavy and show that inflation with the Higgs field can be accommodated in
this framework.

\end{abstract}

\end{titlepage}

\section{Introduction}
\label{sec:intro}

In~\cite{Karananas:2024xja}, the Weyl-invariant Einstein-Cartan~(EC) gravity
nonminimally coupled to the Standard Model (SM) of particle physics was
built. The purely gravitational part of the proposed theory comprises terms
quadratic in the scalar and pseudoscalar (Holst) curvatures, and apart from
the massless graviton, it only propagates an additional pseudoscalar
particle. The communication with the SM is achieved via ``portals'' involving
all possible terms---consistent with the symmetries---built out of the
Higgs/fermionic fields and geometric data, with the restriction to
contributions at most quadratic in curvature and torsion (see also \cite
{Karananas:2021zkl,Karananas:2021gco,Rigouzzo:2023sbb}).

The resulting theory turned out to have a number of virtues. In addition to
the long-known fact that EC gravity can be derived from gauging the Lorenz
group ~\cite{Utiyama:1956sy,Kibble:1961ba,Sciama:1962}, the new pseudoscalar
particle can assume the role of an axion. Generating a sufficiently strong
coupling to QCD turns out to be non-trivial \cite{Karananas:2025ews}, but the
Weyl-invariant EC theory of \cite{Karananas:2024xja} offers a way for
achieving this and for solving the strong CP problem with an axion-like
particle (ALP) of gravitational origin.\footnote
{See~\cite{Wetterich:1983bi,Bardeen:1995kv,Wetterich:1987fk,Wetterich:1987fm,
Dehnen:1992jc,Wetterich:1994bg,Cervantes-Cota:1995ehs,Foot:2007iy,
Shaposhnikov:2008xb,Shaposhnikov:2008xi,Shaposhnikov:2008ar,
Shaposhnikov:2009nk,GarciaBellido:2011de,Blas:2011ac,GarciaBellido:2012zu,
Bezrukov:2012hx,Monin:2013gea,Tavares:2013dga,Khoze:2013uia,Csaki:2014bua,
Rubio:2014wta,Ghilencea:2015mza,Karam:2015jta,Trashorras:2016azl,
Karananas:2016grc,Ferreira:2016vsc,Karananas:2016kyt,Karam:2016rsz,
Ferreira:2016wem,Ferreira:2016kxi,Ghilencea:2016dsl,Shkerin:2016ssc,
Rubio:2017gty,Tokareva:2017nng,Casas:2017wjh,Ferreira:2018itt,
Ferreira:2018qss,Shaposhnikov:2018jag,Burrage:2018dvt,Gorbunov:2018llf,
Lalak:2018bow,Ghilencea:2018thl,Iosifidis:2018zwo,Casas:2018fum,
Shkerin:2019mmu,Herrero-Valea:2019hde,Ghilencea:2019rqj,Karananas:2019fox,
Ghilencea:2020piz,Rubio:2020zht,Gialamas:2020snr,Karananas:2020qkp,
Ghilencea:2020rxc,Hill:2020oaj,Gialamas:2021enw,Gialamas:2021rpr,
Piani:2022gon,Karananas:2021gco,Gialamas:2022xtt,Karananas:2023zgg}
for various attempts to address the naturalness issues via scale or Weyl
invariance and cosmological implications, as well as \cite{He:2025fij} for a
complementary discussion of generating an axion in EC gravity.}
At the same time, the selfconsistency of the model~\cite
{Karananas:2024hoh,Karananas:2024qrz} requires that it has a nonzero
cosmological constant. Both values of the gravitationally-induced ALP mass
and cosmological constant are directly related to the two couplings of the
gauged Lorentz group. Their smallness, as dictated by phenomenology, can be
rephrased in this context as the fact that gravity is the weakest force~\cite
{Arkani-Hamed:2006emk}. As an added bonus, the classical value of the Higgs
mass is controlled by the exact same coupling that fixes the cosmological
constant, so it is vanishingly small. This makes it in principle a computable
quantity. Its smallness as compared to the Planck mass $M_P$ may be
attributed to its nonperturbative generation via gravi-Higgs instantons~\cite
{Shaposhnikov:2018xkv,Shaposhnikov:2018jag,Shaposhnikov:2020geh,
Karananas:2020qkp}.

Here, we take a step back from requiring that the ALP is the QCD axion and we
study the inflationary dynamics when the field is sufficiently heavy and
decouples during inflation. Concretely, we demonstrate that for particular
choices of the nonminimal couplings of the Higgs field to the gravitational
invariants, it is straightforward to Higgs-inflate the Universe. We present
two examples: one where Higgs inflation in metric formulation of
gravity~\cite{Bezrukov:2007ep} is reproduced, and another where a mechanism
fully analogous to the
$\a$-attractors~\cite{Kallosh:2013lkr,Kallosh:2013yoa,Galante:2014ifa} is
operative. Our results are complementary to and extend the findings
of~\cite{Gialamas:2024iyu}, where inflation in this framework was studied in
detail but for a scalar field not identified with the Higgs.

This article is organized as follows. In Sec.~\ref{sec:WI_EC_gravity}, we
give an overview of the Weyl-invariant EC theory nonminimally coupled to the
SM Higgs field. In Sec.~\ref{sec:Decoupled_ALP}, we focus on the inflationary
dynamics of the Higgs field with a decoupled ALP. Our conclusions can be
found in Sec.~\ref{sec:conclusions}. Appendices~\ref {app:no_slow_roll}
and~\ref{app:decoupled_diagonal} supplement the considerations of
Sec.~\ref{sec:Decoupled_ALP}. In the former, we derive the equations of
motion for the single-field setup and study the evolution of the system on a
Friedmann-Lema\^itre-Robertson-Walker (FLRW) spacetime. In the latter, we
demonstrate how the ALP can be decoupled after the action has been
diagonalized.

\section{Weyl-invariant Einstein-Cartan gravity: 
a recap of the gravitational and scalar sectors}
\label{sec:WI_EC_gravity}

The gravitational and scalar parts of the action we are considering
reads~\cite{Karananas:2024xja,Shaposhnikov:2025znm}\footnote{We have not
included the cross-term $R\tilde R$ in the action, since it can be trivially
accounted for~\cite {Gialamas:2024iyu,Karananas:2025xcv}. Its inclusion is
essential only if one wishes to have the ALP as an
inflaton~\cite{Karananas:2025xcv}.}
\begin{align}
\label{eq:infl_action_full}{}
S &= \int \diff^4 x \sqrt{g} \Bigg[ \f{1}{f^2}R^2 
+\f{1}{\tilde f^2}\tilde R^2+ \f{\x_h h^2}{2} R 
+ \f{\z_h h^2}{2} \tilde R \nonumber \\
&~~\qquad\qquad+ \f{c_{aa}h^2}{2} a_\m a^\m 
+\f{c_{\t\t}h^2}{2} \t_{\m\n\rho}\t^{\m\n\rho} 
+ \f{\tilde c_{\t\t}h^2}{2} E^{\m\n\rho\s}
\t_{\lambda\m\n}\t^\lambda_{\ \rho\s} -\f 1 2\l(D_\m^W h\r)^2 
-\f{\lambda h^4}{4} \Bigg] \ .
\end{align} 
Here $g=-{\rm det}(g_{\m\n})$, and $E^{\m\n\rho\s} = \eps^{\m\n\rho\s}/\sqrt
{g}$ is the densitized totally antisymmetric symbol; $f$ and $\tilde f$ are
the gauge couplings of the Lorentz group; $R$ and $\tilde R$ are the scalar
and pseudoscalar invariants
\be
R = g^{\sigma\n} \d^\m_\rho R^\rho_{~\sigma\m\n} \ ,
~~~\tilde R = E^{\rho\s\m\n}R_{\rho\s\m\n} \ ,
\ee
constructed out of the affine curvature tensor
\be
R^\rho_{~\sigma\m\n} = \p_\m  \G^\rho_{~\n\s} - \p_\n \G^\rho_{~\m\sigma} 
+ \G^\rho_{~\m\lambda} \G^\lambda_{~\n\sigma} 
-\G^\rho_{~\n\lambda} \G^\lambda_{~\m\sigma} \ ,
\ee 
with $\G^\m_{~\n\rho}$ the metric-compatible affine connection; its
antisymmetric part is the torsion tensor
\be
T^\m_{~\n\rho} \equiv \G^\m_{~\n\rho} - \G^\m_{~\rho\n} \ ,
\ee 
which for convenience we have decomposed into its irreducible pieces (vector
$v$, axial vector $a$, reduced tensor $\tau$)~\cite{Karananas:2021zkl}
\be
v^\m = g_{\n\rho}T^{\n\m\rho} \ ,
~~~a^\m = E^{\m\n\rho\s}T_{\n\rho\s} \ ,
~~~\tau_{\m\n\rho} =\f 2 3 T_{\m\n\rho} 
+\f 1 3\l(g_{\m\n}v_\rho-g_{\m\rho}v_\n \r) 
- \f 1 3 (T_{\n\rho\m}-T_{\rho\n\m}) \ ,
\ee
with $g^{\m\rho}\tau_{\m\n\rho}=E^{\m\n\rho\s}\tau_{\n\rho\s} = 0$;
$\x_h,\z_h,c_{aa},c_{\t\t},\tilde c_{\t\t}$ are nonminimal couplings of the
Higgs $h$ (in unitary gauge) to the various gravitational invariants;
finally, $\lambda$ is the Higgs self-coupling, whereas $D_\m^W$ is the
Weyl-covariant derivative~\cite{Karananas:2021gco,Karananas:2024xja}
\be
D_\m ^ W h = \p_\m h +\f{v_\m}{3} h \ .
\ee

The equivalent metric theory of~(\ref{eq:infl_action_full}) was presented
in~\cite{Karananas:2024xja}. For completeness, we briefly discuss how to
obtain it here, too.  First, introduce two auxiliary fields, the
dimension-one $\c$ and dimensionless $\phi$, to recast the action as
\begin{align}
S &=\int \diff^4 x \sqrt{g} \Bigg[ \l(\c^2 + \f{\x_h h^2}{2}\r) R  
+\l(M_P^2 \phi + \f{\z_h h^2}{2}\r)\tilde R -\f{f^2\c^4}{4} 
-\f{\tilde f^2 M_P^4\phi^2}{4}  \nonumber \\
&+ \f{c_{aa}h^2}{2} a_\m a^\m 
+\f{c_{\t\t}h^2}{2} \t_{\m\n\rho}\t^{\m\n\rho} 
+ \f{\tilde c_{\t\t}h^2}{2} E^{\m\n\rho\s}
\t_{\lambda\m\n}\t^\lambda_{\ \rho\s} -\f 1 2\l(D_\m^W h\r)^2 
-\f{\lambda h^4}{4} \Bigg] \ .
\end{align}
Then, it is convenient to gauge-fix the Weyl redundancy by setting
$\c=M_P/\sqrt{2}$, such that the above becomes\footnote{In metric-affine
gravity, where geometry is endowed with non-metricity in addition to
curvature and torsion, a very similar theory can be derived from the
requirement of~\emph{extended projective symmetry}~\cite{Barker:2024dhb} (see
discussion in \cite{Karananas:2024xja}).}
\begin{align}
\label{eq:infl_action_auxiliary}
S &=\int \diff^4 x \sqrt{g} \Bigg[ \f{M_P^2+\x_h h^2}{2} R  +\l(M_P^2 \phi 
+\f{\z_h h^2}{2}\r)\tilde R -\f{f^2 M_P^4}{16} 
-\f{\tilde f^2 M_P^4\phi^2}{16}\nonumber \\
&+ \f{c_{aa}h^2}{2} a_\m a^\m +\f{c_{\t\t}h^2}{2} \t_{\m\n\rho}\t^{\m\n\rho} 
+ \f{\tilde c_{\t\t}h^2}{2} E^{\m\n\rho\s}
\t_{\lambda\m\n}\t^\lambda_{\ \rho\s} -\f 1 2\l(D_\m^W h\r)^2 
-\f{\lambda h^4}{4} \Bigg] \ .
\end{align}

Next, the (connection and thus the) curvatures $R$ and $\tilde R$ are
resolved into Riemannian plus post-Riemannian pieces
\begin{align}
\label{eq:R_resol}
&R = \mathring R + 2\mathring \nabla_\m v^\m -\f 2 3 v_\m v^\m 
+ \f{1}{24} a_\m a^\m +\f 1 2 \t_{\m\n\rho} \t^{\m\n\rho} \ ,\\
\label{eq:tildeR_resol}
&\tilde R = -\mathring \nabla_\m a^\m +\f 2 3 a_\m v^\m 
+\f 1 2 E^{\m\n\rho\s} \t_{\lambda\m\n}\t^\lambda_{\ \rho\s} \ ,
\end{align} 
with $\mathring R$ the metrical Ricci scalar and $\mathring\nabla_\m$ the
torsion-free covariant derivative. Plugging~(\ref{eq:R_resol},\ref
{eq:tildeR_resol}) into~(\ref{eq:infl_action_auxiliary}) yields
\begin{align}
\label{eq:infl_action_auxiliary_2}
S &=\int \diff^4 x \sqrt{g} \Bigg[ \f{M_P^2+\x_h h^2}{2} \mathring R 
-\f 1 2 (\p_\m h)^2 -\f{\lambda h^4}{4} -\f{f^2 M_P^4}{16} 
-\f{\tilde f^2M_P^4 \phi^2}{16} \nonumber \\
&+\f{c'_{vv}}{2}v_\m v^\m + \f{c'_{aa}}{2} a_\m a^\m +c'_{va}a_\m v^\m 
+\f{c'_{\t\t}}{2} \t_{\m\n\rho}\t^{\m\n\rho} 
+ \f{\tilde c'_{\t\t}}{2} E^{\m\n\rho\s}\t_{\lambda\m\n}\t^\lambda_{\ \rho\s}
+ v^\m J^v_\m + a^\m J^a_\m \Bigg] \ ,
\end{align}
where we integrated some terms by parts, and defined
\begin{align}
&c'_{vv} = -\f 2 3\l(M_P^2+\l(\x_h +\f 1 6\r) h^2\r),
~c'_{aa} = \f{M_P^2}{24}+\l(\f{\x_h}{24} + c_{aa}\r)h^2,
~c'_{va}= \f 2 3 \l(M_P^2\phi+\f{\z_h h^2}{2}\r),\nonumber \\
&c'_{\t\t} = \f{M_P^2+\l(\x_h+2c_{\t\t}\r) h^2}{2} \ ,
~~~\tilde c'_{\t\t} = M_P^2\phi + \f{\z_h+\tilde c_{\t\t}}{2}h^2 \ ,
\end{align}
and 
\be
J^v_\m = -\l(\x_h +\f 1 6\r)\p_\m h^2\ ,
~~~J^a_\m = \p_\m \l(M_P^2 \phi +\f{\z_h}{2} h^2\r) \ . 
\ee

Now, we vary the action wrt to $v,a,\t$ to obtain the torsional equations of
motion
\be
v_\m = - \f{c'_{aa} J^v_\m - c'_{va} J^a_\m}{c'_{vv}c'_{aa}-c_{va}^{'2}}\ ,
~~~a_\m = - \f{c'_{vv}J^a_\m - c'_{va} J^v_\m}{c'_{vv}c'_{aa}-c_{va}^{'2}}\ ,
~~~\tau_{\m\n\rho} = 0 \ .
\ee
Upon plugging these back in the action~(\ref{eq:infl_action_auxiliary_2}), we
obtain
\be
\label{eq:Jordan_frame_action}
S=\int \diff^4 x \sqrt{g} \Bigg[\f{M_P^2+\x_h h^2}{2}\mathring R 
-\f 1 2\g_{ab}\p_\m \varphi^a\p^\m\varphi^b -\f{\lambda h^4}{4}
-\f{\tilde f^2 M_P^4 \phi^2}{4} - \f{f^2 M_P^4}{16}  \Bigg] \ , 
\ee 
where summation over all repeated indexes is understood. We introduced the
shorthand notation $\varphi^a=(h,\phi)$, and $\g_{ab}=\g_{ab}(h,\phi)$ is the
metric of the two-dimensional kinetic manifold with components
\be
\label{eq:metric_components}
\g_{hh} =\f{N}{D} \ ,
~~~\g_{h\phi} = \f{48\l(3\z_h -(1+6\x_h)\phi\r)M_P^4 h}{D} \ ,
~~~\g_{\phi\phi} = \f{24\l(6M_P^2+(1+6\x_h)h^2\r)M_P^4}{D}\ ,
\ee
where 
\begin{align}
&N= 6(1+16\phi^2)M_P^4+36
\l(4(\z_h^2+c_{aa})-\x_h^2 -16\x_h \z_h \phi\r)M_P^2h^2\nonumber \\
&\qquad\qquad\qquad\qquad\qquad
-6\x_h\l(24\z_h^2+(\x_h+24c_{aa})(1+6\x_h)\r)h^4 \ , \\
\label{eq:denominator}
&D = 6(1+16\phi^2)M_P^4 
+\l(1+12\x_h+ 144c_{aa}+96\z_h \phi\r)M_P^2 h^2\nonumber\\
&\qquad\qquad\qquad\qquad\qquad\qquad
+\l(24\z_h^2 +(\x_h+24c_{aa})(1+6\x_h)\r)h^4 \ .
\end{align}

The last step consists in Weyl-rescaling the metric
\be
\label{eq:weyl_rescaling}
g_{\m\n}~\mapsto~\Omega^{-2} g_{\m\n} \ ,
~~~\Omega^2 = \f{M_P^2+\x_h h^2}{M_P^2} \ ,
\ee
so that the theory is expressed in the Einstein frame: 
\be
\label{eq:action_higgs_Einstein}
S_{\rm E} = \int \diff^4x \sqrt{g} \Bigg[\f {M_P^2}{2}\mathring R 
-\f 1 2\bar\g_{ab}\p_\m \varphi^a\p^\m\varphi^b 
-\f{M_P^4}{4(M_P^2+\x_h h^2)^2}
\l(\lambda h^4+\tilde f^2M_P^4\phi^2+ \f{f^2 M_P^4}{4}\r) \Bigg] \ .
\ee 
The components of the Weyl-transformed field-space metric $\bar\g_{ab}$ are
related to~(\ref{eq:metric_components}) as
\be
\label{eq:metric_components_Einstein}
\bar \g _{hh}= \f{M_P^2}{M_P^2+\x_h h^2}
\l(\g_{hh} +\f{6\x_h^2 h^2}{M_P^2+\x_h h^2}\r) \ ,
~~~\bar \g_{h\phi} = \f{M_P^2\g_{h\phi}}{M_P^2+\x_h h^2} \ ,
~~~\bar \g_{\phi\phi}= \f{M_P^2\g_{\phi\phi}}{M_P^2+\x_h h^2} \ .
\ee
It can be easily checked that the field-space metric is nondegenerate and of
rank 2, therefore the action describes two scalars minimally coupled to
gravity but with highly nontrivial kinetic (and potential) terms, aftermath
of the graviscalar mixings. The theory \eqref{eq:action_higgs_Einstein}
selects a particular subclass of Higgs inflation scenarios within EC
gravity~\cite{Bauer:2008zj,Langvik:2020nrs, Shaposhnikov:2020gts} and
beyond~\cite{Rasanen:2018ihz,Raatikainen:2019qey,Rigouzzo:2022yan}.

\section{Decoupled ALP}
\label{sec:Decoupled_ALP}

Let us assume that the ALP is sufficiently heavy\footnote{Specifically, its
mass should be bigger than the inflationary Hubble scale.} and thus
decouples. This corresponds to formally taking the limit $\tilde f\to\infty$
in the original action~(\ref{eq:infl_action_full})---equivalently, one can
set $\phi=0$ in~(\ref{eq:action_higgs_Einstein}), which is dictated from the
ALP equations of motion.\footnote{In Appendix~\ref {app:decoupled_diagonal},
we diagonalize the kinetic sector of the
action~(\ref{eq:action_higgs_Einstein}). Exactly the same conclusions are
reached if one opts for working with the fields in \eqs (\ref
{eq:canonical_axion_higgs_2},\ref{eq:canonical_ALP_higgs_2}).}

Generally speaking, one can relax the requirement that $\tilde f$ is large.
In this case, to decouple the ALP, the field would need to acquire a large
induced mass from a different source---this can for instance happen if QCD
confines at early times~\cite{Dvali:1995ce}. Although somewhat speculative,
this is an interesting idea: it would mean that $\tilde f$ can be tiny so the
gravitational solution to the strong-CP problem persists.

In any event and irrespectively of the details, with a decoupled ALP the
early Universe dynamics simplifies greatly: $\phi$ does not play any role in
inflation and the theory is effectively single-field.\footnote{In~\cite
{Gialamas:2024iyu}, the single-field regime of the theory was studied, but
without identifying $h$ with the SM Higgs. This analysis uncovered viable
inflation for a wide range of parameters. However, neither the
Higgs-inflationary nor the $\a$-attractors subclasses were found in the
aforementioned article. For further inflationary models based on scale- and
Weyl-invariance, we refer the reader to
\cite{Ghilencea:2019rqj,Ghilencea:2020rxc,Ghilencea:2020piz,Barker:2024goa,
Karananas:2025xcv}.}
At the same time, the resulting dynamics is general enough to
accommodate~(metrical) Higgs inflation, as well as $\a$-attractors. We shall
make that manifest now.

The tree-level Einstein-frame action reads
\be
\label{eq:action_single-field}
S_{\rm E} =\int \diff^4x \sqrt{g} \Bigg[\f{M_P^2}{2}\mathring R 
-\f 1 2 K(h)(\p_\m h)^2 - V(h) -\f{f^2M_P^4}{16(M_P^2+\x_h h^2)^2} \Bigg] \ ,
\ee
with
\be
\label{eq:potE}
V(h) = \f{\lambda h^4 M_P^4}{4(M_P^2+\x_h h^2)^2} \ . 
\ee 

As discussed in~\cite{Karananas:2024xja}, the phenomenologically interesting
situation requires a vanishingly small cosmological constant (controlled by
$f$) meaning that $f\lll \lambda<1$; and with the ALP out of the picture, we
need only retain the (standard) $\lambda$-piece~(\ref{eq:potE}) of the
potential for the Higgs in what follows.

To keep the expression~(\ref{eq:action_single-field}) short, we introduced
the kinetic function
\begin{align}
\label{eq:kinetic_function_Higgs_noALP}
K(h) = &\f{6M_P^6}{(M_P^2+\x_h h^2)^2}\times\nonumber\\
&\qquad\times\f{M_P^2+\l(\x_h+24(c_{aa}+\z_h^2)\r)h^2}
{6M_P^4+(1+12\x_h+144c_{aa})M_P^2h^2
+\l(24\z_h^2+(\x_h+24c_{aa})(1+6\x_h)\r)h^4} \ .
\end{align}

A few comments are in order concerning the above. First, at small Higgs
values relevant for the present-day dynamics, we observe that
\be
K(h) \sim 1 +\mathcal O(h^2) \ ,
\ee
guaranteeing that the field has a canonical kinetic term, irrespectively of
the values of the three parameters ($\x_h,\z_h,c_{aa}$) appearing
in~(\ref{eq:kinetic_function_Higgs_noALP}).

Second, the form of the kinetic function for large Higgs values relevant for
the inflationary dynamics~\emph{crucially depends on $c_{aa}$ and $\z_h$}.
For instance, provided that the parameters are completely unconstrained, the
kinetic function  asymptotes to
\be
K(h) \sim \f{1}{h^6} \ ,
\ee 
when $h$ dominates. Such a sextic ``pole'' translates into a very steep
potential incapable of supporting slow-roll for large field values. This can
be made maximally explicit by computing the slow-roll parameters,\footnote
{This can also be seen by writing down the equations of motion for the
homogeneous Higgs on top of an FLRW spacetime; see Appendix~\ref
{app:no_slow_roll}.} which for a non-canonical field (see e.g.~\cite
{Casas:2018fum})) read
\be
\label{eq:SR_defs}
\eps(h) = \f{M_P^2}{2K(h)}\l(\f{V'(h)}{V(h)}\r)^2 \ ,
~~~\eta(h) = \f{M_P^2}{\sqrt{K(h)}V(h)}\l(\f{V'(h)}{\sqrt{K(h)}}\r)' \ ,
\ee
where a prime stands for differentiation wrt $h$.\footnote{The second
slow-roll parameter can be equivalently expressed as
\be
\label{eq:eta_equiv_form}
\eta(h) = 2\eps(h) + \eps'(h) \f{V(h)}{V'(h)} \ .
\ee} Using~(\ref{eq:potE},\ref{eq:kinetic_function_Higgs_noALP}), the above
 yield
\begin{align}
\label{eq:SR_eps_expanded}
&\eps(h) = \f{4}{3}\l(1 +6\x_h -(12\z_h)^2 +\f{6M_P^2}{h^2}
+\f{24(12\z_h)^2(c_{aa}+\z_h^2)h^2}{M_P^2
+\l(\x_h+24(c_{aa}+\z_h^2)\r)h^2}\r) \ ,\\
\label{eq:SR_eta_expanded}
&\eta(h) = 2\l(\eps - 2\x_h -\f{2M_P^2}{h^2}
+\f{8(12\z_h)^2(c_{aa}+\z_h^2)h^2
\l(M_P^2+\x_h h^2\r)}{\l(M_P^2+\l(\x_h+24(c_{aa}+\z_h^2)\r)h^2\r)^2}\r) \ .
\end{align}
It is obvious that for large $h$, the slow-roll parameters become constant
\begin{align}
&\eps \sim \f{4}{3}\l(1 +6\x_h -(12\z_h)^2
+\f{24(12\z_h)^2(c_{aa}+\z_h^2)}{\x_h+24(c_{aa}+\z_h^2)}\r) \ ,\\
&\eta \sim \f 8 3\l(1 +\f{3\x_h}{\x_h+24(c_{aa}+\z_h^2)}
\l(\x_h+36c_{aa}+\f{\x_h(\x_h+24c_{aa})}
{2\l(\x_h+24(c_{aa}+\z_h^2)\r)}\r)\r) \ ,
\end{align} 
from which we conclude that a sufficient period of slow-roll inflation
($\eps,\eta\ll 1$) may only be achieved for specific choices of $c_{aa}$ and
$\z_h$. In other words, one needs to ``shape'' the large-field behavior of
the kinetic function $K(h)$.

It turns out that not only is this possible to do, but also the
Weyl-invariant EC gravity with a heavy ALP comprises subclasses with
the~\emph{Higgs as inflaton} which are well-known to be in excellent
agreement with the Planck/BICEP results~\cite{Planck:2018jri,BICEP:2021xfz}.
As for a possible tension with
ACT/DESI~\cite{ACT:2025fju,ACT:2025tim,DESI:2024uvr,DESI:2024mwx}, we prefer
to wait until the dust settles; see discussion in \cite{Ferreira:2025lrd}.

\subsection{Metric Higgs inflation} 

One option is to require that the coefficients multiplying the quadratic and
quartic powers of $h$ in the denominator of the second line in expression
(\ref{eq:kinetic_function_Higgs_noALP}) vanish. This is achieved by setting
\be
\label{eq:parameters_Higgs_infl_noALP}
c_{aa} = \f{1}{144}\l(1\mp 24\z_h\r) \ ,
~~~\z_h = \pm \f{1+6\x_h}{12} \ ,
\ee
which in turn results into
\be
\label{eq:Kfunction_HI}
K(h) = \f{M_P^2\l(M_P^2+\x_h(1+6\x_h)h^2\r)}
{\l (M_P^2 + \x_h h^2\r)^2} \ ,
\ee 
with the primeval inflationary epoch corresponding to 
\be
\label{eq:higgs_infl_dom}
h\gg \f{M_P}{\sqrt{\x_h}} \ .
\ee

Remarkably, the expression~(\ref{eq:Kfunction_HI}) is \emph{exactly the
kinetic function of the usual Higgs inflation in metric gravity}. Therefore,
it should hardly come as a surprise that the single-field counterpart
of~(\ref{eq:Jordan_frame_action}) evaluated
on~(\ref{eq:parameters_Higgs_infl_noALP}), is found to be
\be
S = \int\diff^4 \sqrt{g} \Bigg[\f{M_P^2+\x_h h^2}{2}\mathring R 
- \f 1 2 (\p_\m h)^2 -\f{\lambda}{4} h^4 - \f{f^2M_P^4}{16}\Bigg] \ ,
\ee 
which is the vanilla Higgs inflation action~\cite{Bezrukov:2007ep} in the
Jordan frame. 

The slow-roll parameters for $c_{aa}$ and $\z_h$ given
by~(\ref{eq:parameters_Higgs_infl_noALP}), are
\begin{align}
&\eps(h) = \f{8M_P^4}{h^2\l(M_P^2+\x_h(1+6\x_h)h^2\r)} \ ,\\
&\eta(h) = \f{4M_P^2\l(3M_P^4+\x_h h^2\l(M_P^2(1+12\x_h)
-2(1+6\x_h)\x_h h^2\r)\r)}{h^2\l(M_P^2+\x_h(1+6\x_h)h^2\r)^2} \ ,
\end{align}
which for~(\ref{eq:higgs_infl_dom}) boil down to 
\be
\eps(h) \sim \f{8M_P^4}{(1+6\x_h) \x_h h^4} \ll 1 \ ,
~~~|\eta(h)| \sim \f{8M_P^2}{(1+6\x_h)h^2} \ll 1 \ ,
\ee
as they should. 

CMB normalization at horizon exit 
\be
\label{eq:CMB_normal}
\frac{V(h)}{\eps(h)} = 5\times 10^{-7}~M_P^4 \ , 
\ee 
dictates that, for the typical SM value of the Higgs self-coupling
$\lambda\sim \mathcal O(10^{-2})$, the nonminimal coupling is 
\be
\x_h\sim \mc O\l(10^{3-4}\r) \ .
\ee
The tilt $n_s$ and tensor-to-scalar ratio $r$ are well-known 
\be
\label{eq:indexes_HI}
n_s\approx 1 -\f{2}{N} \ ,~~~r \approx \f{12}{N^2} \ ,
\ee 
where $N$ is the number of inflationary e-foldings given by
\be \label{Nint}
N = \f{1}{M_P}\limitint^{h_*}_{h_{\rm e}} 
\diff h\sqrt{\f{K(h)}{2\eps(h)}} \ ,
\ee
with $h_*$ and $h_{\rm e}$ the field values at horizon exit and end of
inflation, respectively.

\subsection{``Higgs $\alpha$-attractors''}

Yet another interesting option is to arrange for the kinetic function to
exhibit a quadratic pole at some field value. Indeed, setting
\be
\label{eq:parameters_attractor_infl_noALP}
c_{aa} = \f{1}{144}\l(1\mp 24\z_h\r) \ ,
~~~\z_h=\pm \f{1}{12}\l(1+6\a+6\x_h\r)\ ,~~~\a > 0 \ ,
\ee
we find that
\be 
\label{eq:Kfunction_attractors}
K(h) = \f{M_P^6}{(M_P^2+\x_h h^2)^2}
\f{M_P^2+\l(\x_h+6(\a+\x_h)^2\r)h^2}{(M_P^2-\a h^2)^2} \ ,
\ee 
with inflation in this setup materializing at 
\be
\label{eq:attractor_dom}
h\mapsto\f{M_P}{\sqrt{\a}} \ .
\ee

There is no difficulty in computing the $\eps$ and $\eta$ parameters
\begin{align}
&\eps(h) = \f{8\l(M_P^2-\a h^2\r)^2}
{h^2\l(M_P^2+\l(\x_h+6(\a+\x_h)^2\r)h^2\r)} \ , \\ 
&\eta(h) = \f{4\l(M_P^2-\a h^2\r)}
{h^2\l(M_P^2+\l(\x_h+6(\a+\x_h)^2\r)h^2\r)^2}\times\\
&\qquad\qquad\times\Bigg[3M_P^4+(\a+\x_h)
\l(12(\a+\x_h)+1-\f{6\a}{\a+\x_h}\r)M_P^2h^2\\
&\qquad\qquad\qquad-(\a+\x_h)
\l(12(\a+\x_h)^2+2(\a+\x_h)(1+6\a)+\a\l(1-\f{3\a}{\a+\x_h}\r)\r)h^4\Bigg] \ , 
\end{align}
which are immediately seen to vanish for~(\ref{eq:attractor_dom}),
guaranteeing that slow-roll is possible.

We now expand~(\ref{eq:Kfunction_attractors}) around the inflationary pole,
which gives
\be \label{KAlphaAttractor}
K(h) \approx \f{3 M_P^2}{2 \l(\f{M_P}{\sqrt{\a}}-h\r)^2}
\l(1 - \f{\sqrt{\a} \l(\a-3\x_h\r)\l(\f{M_P}{\sqrt{\a}}-h\r)}
{\l(\a+\x_h\r) M_P } +\ldots \r) \ ,
\ee 
with ellipses standing for finite and subleading terms. Moreover, we already
used that successful inflation requires $\a+\x_h\gg 1$, as we shall show
shortly. At the same time, the potential is well approximated by
\be
V(h) \approx \f{\lambda M_P^4}{4\l(\a+\x_h\r)^2}
\l(1- \f{4\a^{3/2}\l(\f{M_P}{\sqrt{\a}}-h\r)}{M_P(\a+\x_h)} +\ldots\r) \ .
\ee

With these approximations and only using the leading terms, we can compute
\be
\eps(h) \approx \f{16 \a^3 \l(\f{M_P}{\sqrt{\a}}-h\r)^2}
{3 \l(\a+\x_h\r)^2 M_P^2} \;,
\ee
so that \eq \eqref{Nint} yields
\be
\f{M_P}{\sqrt{\a}}-h \approx \f{3 M_P \l(\a+\x_h\r)}{8 \a^{3/2} N} \ .
\ee
Plugging this into \eq \eqref{KAlphaAttractor}, we see that restricting
ourselves to leading terms is only viable as long as $\x_h \ll \a N$. With
$h$ as a function of $N$, we can evaluate the amplitude of perturbations
observed in the CMB from~(\ref{eq:CMB_normal}) to conclude that
\be
\f{\lambda}{\l(\a+\x_h\r)^2} \approx \f{1.5\times 10^{-6}}{N^2} \ .
\ee
Finally, the inflationary indices are given by
\be
\label{eq:indexes_attractors}
n_s \approx 1 -\f{2}{N} \ ,~~~r \approx \f{12}{N^2} \ ,
\ee
in accordance with the standard expressions~\cite{Galante:2014ifa} for
$\a$-attractors. Remarkably, these observables are independent of both $\a$
and $\x_h$ and so indistinguishable from the outcome of metric Higgs
inflation as shown in \eq \eqref{eq:indexes_HI}. Thus, generalizing the
parameter choice \eqref{eq:parameters_Higgs_infl_noALP} by introducing $\a$
as in \eq \eqref{eq:parameters_attractor_infl_noALP} does not influence
inflationary observables at leading order.

\section{Conclusions}
\label{sec:conclusions}

With a heavy axion-like particle, the Weyl-invariant theory encompassing
Einstein-Cartan gravity and the Standard Model of particle
physics~\cite{Karananas:2024xja} is capable of Higgs-inflating the Universe
in different ways and with predictions in excellent agreement with the
Planck/BICEP cosmological data.

\section*{Acknowledgements}

The work of M.S.~was supported in part by the Generalitat Valenciana grant
PROMETEO/2021/083. S.Z.~acknowledges support by the European Research Council
Gravites Horizon Grant AO number: 850 173-6.

\leavevmode

\textbf{Disclaimer.} Funded by the European Union. Views and opinions
expressed are however those of the authors only and do not necessarily
reflect those of the European Union or European Research Council. Neither the
European Union nor the granting authority can be held responsible for them.

\appendices

\section{Cosmological evolution with a decoupled ALP}
\label{app:no_slow_roll}

Here we discuss the background dynamics of the action of Sec.~\ref
{sec:Decoupled_ALP}; we follow the corresponding discussion of~\cite
{Garcia-Bellido:2011kqb} and the Appendix B of~\cite{Karananas:2024xja}, but
adapted to the single-field case. Consider a flat
Friedmann-Lema\^itre-Robertson-Walker(FLRW) spacetime with metric
\be
\label{eq:FLRW_metric}
g_{\m\n} = {\rm diag}(-1,a^2,a^2,a^2) \ ,~~~a\equiv a(t) \ ,
\ee 
and take the Higgs to be homogeneous, $h=h(t)$. The equations of motion
following from~(\ref{eq:action_single-field}) on the background~(\ref
{eq:FLRW_metric}) are
\bea
&&3M_P^2 \mc H^2 = \f{K(h)}{2}\dot h^2 + V(h) \ ,\\
&&M_P^2\l(2\dot {\mc H} + 3\mc H^2\r) = -\f{K(h)}{2}\dot h^2 + V(h) \ ,\\
&&K(h)\l(\ddot h + 3 \mc H \dot h \r)+\f{K'(h)}{2}\dot h^2 = -V'(h) \ ,
\eea
where $\cdot$ and $'$ stand for differentiation wrt $t$ and $h$,
respectively. The Hubble parameter is defined as $\mc H=\dot a/a$.

In terms of the inflationary e-folding parameter $N\equiv\log a$, the above
are recast as
\bea
\label{eq:app_eom1}
&& \mc H^2 = \f{V(h)}{3M_P^2-\f{K(h)}{2}\l(\f{dh}{dN}\r)^2} \ ,\\
&& \f{1}{\mc H} \f{d\mc H}{dN} = -\f{K(h)}{2M_P^2}\l(\f{dh}{dN}\r)^2 \ ,\\
\label{eq:app_eom2}
&& \f{\f{d^2h}{dN^2}+\f{K'(h)}{2K(h)}\l(\f{dh}{dN}\r)^2}
{3-\f{K(h)}{2M_P^2}\l(\f{dh}{dN}\r)^2}+\f{dh}{dN} 
= -\f{M_P^2}{K(h)}\f{V'(h)}{V(h)} \ .
\eea

The ``cosmological'' slow-roll parameters $\eps_{\mc H}$ and $\eta_{\mc H}$
are defined as
\be
\eps_{\mc H} = -\f{1}{\mc H}\f{d\mc H}{dN} 
= \f{K(h)}{2M_P^2}\l(\f{dh}{dN}\r)^2 \ ,
~~~\eta_{\mc H} = 
\f{\f{d^2h}{dN^2}+\f{K'(h)}{2K(h)}\l(\f{dh}{dN}\r)^2}{\f{dh}{dN}} \ ,
\ee
respectively. The equations~(\ref{eq:app_eom1},\ref{eq:app_eom2}) therefore
become
\bea
&&M_P^2 \mc H^2 = \f{V(h)}{3-\eps_{\mc H}} \ ,\\
\label{eq:app_eomh}
&&\f{dh}{dN} =-\f{M_P^2}{K(h)}\f{V'(h)}{V(h)}
\l(1-\f{\eta_{\mc H}}{3-\eps_{\mc H}+\eta_{\mc H}}\r) \ .
\eea

Accelerated expansion for sufficient amount of time corresponds to $\eps_
{\mc H},~|\eta_{\mc H}| <1$, which means that the system in this slow-roll
phase is well approximated by
\be
M_P^2 \mc H^2 \approx \f{V(h)}{3} \ ,
~~~\f{dh}{dN} \approx-\f{M_P^2}{K(h)}\f{V'(h)}{V(h)} \ .
\ee 
Upon plugging~(\ref{eq:potE},\ref{eq:kinetic_function_Higgs_noALP}) into the
above, we find
\begin{align}
\f{dh}{dN} \approx &- \f{2(M_P^2+\x_h h^2)}{3M_P^2 h}\times\nonumber\\
 &\qquad\times\f{6M_P^4+(1+12\x_h+144c_{aa})M_P^2h^2
 +\l(24\z_h^2+(\x_h+24c_{aa})(1+6\x_h)\r)h^4}
 {M_P^2+\l(\x_h+24(c_{aa}+\z_h^2)\r)h^2} \ .
\end{align} 
As concluded in the main text (Sec.~\ref{sec:Decoupled_ALP}), unless the
nonminimal couplings $c_{aa}$ and $\z_h$ are carefully chosen, the right-hand
side of the above grows instead of decaying at large field values. Therefore,
there is no viable inflation.

\section{Decoupled ALP in diagonalized action}
\label{app:decoupled_diagonal}

There exists a change of variables such that the kinetic sector of~(\ref
{eq:action_higgs_Einstein}) can be fully diagonalized and with a canonical
Higgs~\cite{Karananas:2024xja}. Indeed, introducing the fields $H$ and
$\Phi$, related to $h$ and $\phi$ as
\begin{align}
\label{eq:canonical_axion_higgs_2}
&H = \sqrt 6 M_P\, \arctanh \l(\f{\f{h}{M_P}}
{\sqrt{6+(1+6\x_h)\f{h^2}{M_P^2}}}\r) \ ,\\
\label{eq:canonical_ALP_higgs_2}
&\Phi = \f{M_P}{1+6\x_h}\log\l(\f{6+(1+6\x_h)\f{h^2}{M_P^2}}
{\l|3\z_h-(1+6\x_h)\phi\r|}\r) \ ,
\end{align}
one finds that the action~(\ref{eq:action_higgs_Einstein}) becomes
\be
\label{eq:canonical_Higgs_action}
S_{\rm E} = \int\diff^4x\sqrt{g}\Bigg[\f{M_P^2}{2} \mathring R 
-\f 1 2(\p_\m H)^2 -\bar\g_{\Phi\Phi}(\p_\m\Phi)^2 -V(H,\Phi) \Bigg] \ . 
\ee
The kinetic function for the ALP is
\be
\bar\g_{\Phi\Phi} = \f 3 4 \f{\cosh^4\l(\f{H}{\sqrt 6 M_P}\r)}
{\cosh^2\l(\f{H}{\sqrt 6 M_P}\r)
\l(\f{1-\f{\z_h}{2}e^{(1+6\x_h)\f{\Phi}{M_P}}}{1+6\x_h}\r)^2
+\l(\f{e^{(1+6\x_h)\f{\Phi}{M_P}}}{24}\r)^2
\l(1+144c_{aa}\sinh^2\l(\f{H}{\sqrt 6 M_P}\r)\r)} \ ,
\ee
whereas the potential reads
\be
\label{eq:potential_canonical_fields}
V(H,\Phi)=  V_\lambda(H) + V_f(H)+V_{\tilde f}(H,\Phi)\ ,
\ee
with
\begin{align}
\label{eq:Vlambda}
&V_\lambda(H) = 9\lambda M_P^4\sinh^4\l(\f{H}{\sqrt 6 M_P}\r) \ ,\\
\label{eq:Vf}
&V_f(H) = \f{f^2M_P^4}{16}\l(1-6\x_h\sinh^2\l(\f{H}{\sqrt 6 M_P}\r)\r)^2 \ ,\\
\label{eq:Vtildef}
&V_{\tilde f}(H,\Phi) = \f{9\tilde f^2 M_P^4 e^{-2(1+6\x_h)\f{\Phi}{M_P}}}
{(1+6\x_h)^2}\Bigg[\cosh^2\l(\f{H}{\sqrt 6 M_P}\r)\nonumber\\
&\qquad\qquad\qquad-\f{\z_h}{2} e^{(1+6\x_h)\f{\Phi}{M_P}}
\l(1-6\x_h\sinh^2\l(\f{H}{\sqrt 6 M_P}\r)\r)\Bigg]^2 \ .
\end{align}

It is of course possible to decouple the ALP also when working in terms of
the diagonal variables $H$ and $\Phi$. The minimum of the above in the
$\Phi$-direction is located at
\be
\label{eq:flat}
\Phi = \f{M_P}{1+6\x_h} \log\l[\f{2\cosh^2\l(\f{H}{\sqrt 6 M_P}\r)}
{\z_h\l(1-6\x_h \sinh^2\l(\f{H}{\sqrt 6 M_P}\r)\r)}\r] \ .
\ee 
As expected, this relation can also be obtained by plugging~(\ref
{eq:canonical_axion_higgs_2}) into~(\ref{eq:canonical_ALP_higgs_2}) and
setting $\phi=0$.

The action~(\ref{eq:canonical_Higgs_action}) evaluated on~(\ref {eq:flat}),
becomes
\begin{align}
\label{eq:app_single_field_H}
S= \int\diff^4x\sqrt{g} \Bigg[&\f{M_P^2}{2}\mathring R 
- \f{\l(1+144(\z_h^2+c_{aa})\sinh^2\l(\f{H}{\sqrt 6 M_P}\r)\r)
\cosh^2\l(\f{H}{\sqrt 6 M_P}\r)}{1+72c_{aa}\sinh^2\l(\sqrt{\f 2 3}
\f{H}{ M_P}\r)+288\z_h^2\sinh^4\l(\f{H}{\sqrt 6 M_P}\r)}(\p_\m H)^2\nonumber\\
&\quad -9\lambda M_P^4\sinh^4\l(\f{H}{\sqrt 6 M_P}\r)-\f{f^2M_P^4}{16}
\l(1-6\x_h\sinh^2\l(\f{H}{\sqrt 6 M_P}\r)\r)^2 \Bigg] \ ,
\end{align} 
where the piece of the potential proportional to $\tilde f$ vanishes along
the flat direction~(\ref{eq:flat}). The above is exactly~(\ref
{eq:action_single-field}), as it can be straightforwardly verified by using
the transformation~(\ref{eq:canonical_axion_higgs_2}) to express~(\ref
{eq:app_single_field_H}) in terms of $h$.

{
\setlength\bibsep{0pt}
    \bibliographystyle{utphys}
 \bibliography{Refs}
}

\end{document}